# On the possibility of superluminal energy propagation in a hyperbolic metamaterial of metal-dielectric layers

Pi-Gang Luan[1] and Jie-Luen Wu[2]

[1] *Department of Optics and Photonics, National Central University, Jhongli District, Taoyuan City 320, Taiwan (R.O.C)*

[2] *System Development Center, National Chung-Shan Institute of Science and Technology, Longtan Township, Taoyuan City 325, Taiwan (R.O.C)*

The energy propagation of electromagnetic fields in the effective medium of a one-dimensional photonic crystal consisting of dielectric and metallic layers is investigated. We show that the medium behaves like Drude and Lorentz medium, respectively, when the electric field is parallel and perpendicular to the layers. For arbitrary time-varying electromagnetic fields in this medium, the energy density formula is derived. We prove rigorously that the group velocity of any propagating mode obeying the hyperbolic dispersion must be slower than the speed of light in vacuum, taking into account the frequency dependence of the permittivity tensor. That is, it is not possible to have superluminal propagation in this dispersive hyperbolic medium consisting of real dielectric and metallic material layers. The propagation velocity of a wave packet is also studied numerically. This packet velocity is very close to the velocity of the propagating mode having the central frequency and central wave vector of the wave packet. When the frequency spread of the wave packet is not narrow enough, small discrepancy between these two velocities manifests, which is caused by the non-penetration effect of the evanescent modes. This work reveals that no superluminal phenomenon can happen in a dispersive anisotropic metamaterial medium made of real materials.

**I. INTRODUCTION**

Metamaterials usually refer to artificially designed structures consisting of periodically arranged resonators [1], dielectric-metal layers [2] or intrinsic resonant components for achieving unusual electromagnetic properties such as negative refractive index or indefinite permittivity [3,4]. With such unusual properties, novel devices such as superlens [5] and hyperlens [6] for subwavelength imaging and invisibility cloak [7] for hiding objects can be realized experimentally in practice. The resonance and directional-conductive characters of these metamaterials imply that they are inherently dispersive, absorptive, and usually anisotropic. A great many research works have already been done for understanding the propagating properties of electromagnetic waves in these media, and various novel devices have been proposed and tested [8-12]. However, studies concerning how wave packets or signals propagating in these dispersive media are few [13]. The wave packet propagation problem is theoretically interesting because it is related to some fundamental problems like the causality or superluminality. It may also be technically relevant because with the knowledge about how wave packets propagate and distort in the media, we may develop useful techniques for controlling or manipulating the signals evolving in these media, which is highly desirable for applications related to time dependent phenomena.

It is well known that any kind of wave propagation process must satisfy causality [14]. That is, the cause must happen before the appearance of the result. Special relativity tells us that if superluminal signals can be sent as messages connecting event A and event B in an inertial frame, then it is always possible to find an inertial frame in which the observed orders of time for them are exchanged, violating the constraint of causality. Another restriction deduced from relativity is that no massive particle can move faster than the speed of light in a vacuum [15]. These considerations imply that waves can never have superluminal propagation velocity for the energy or information carried by them. However, recent studies concerning the propagation of electromagnetic [16], acoustic [17], and quantum waves [18] in various kinds of dispersive and dissipative media revealed that, under suitable conditions and through various mechanisms, superluminal phenomena do exist and can be observed, though in some subtle ways they never break causality and do not violate the restriction of speed limitation from relativity [19].

Up to now, most studies about the superluminal phenomena considered isotropic media or one-dimensional propagation [20-26]. About a decade ago, a study claimed that the group velocity of hyperbolic propagating mode can become superluminal under suitable conditions [27]. However, a recent theoretical work concerning passive linear media with negligible loss proved that the group/energy velocity of the electromagnetic propagating modes can never exceed the speed of light in a vacuum [28]. These contradictive results attracted our attention and inspired us to further explore the propagation behaviors of wave energy in a dispersive-anisotropic medium.

In this paper, we provide a rigorous proof showing that the group velocity of any propagating mode inside the effective medium that derived from the periodic structure of dielectric and metallic layers, no matter whether it obeys elliptical or hyperbolic dispersion, is always slower than the speed of light in a vacuum. This is the most important result of this paper. We also numerically simulate the wave packet propagating behavior based on the Fourier transform method and the energy density formula derived in this paper. Although our result on the mode propagating velocity confirmed the conclusion given in [28], our proof is concrete in details and directly based on the material properties, different from that of the more abstract approach in [28]. Besides, though in the lossless limit our time dependent energy density formula applying to a propagating mode is the same as that derived from the corresponding formula in [28], for the case containing nonzero loss our energy density formula follows our previous work [29], excluding the "dissipation part" in the formula of [28] that represents the time integration of the power loss. In addition, very detailed results about the signatures and dispersions of the principal values of permittivity tensor are given, which is rare in existing publications.

The metamaterials discussed in this paper can be realized effectively in the long-wavelength limit by the one-dimensional photonic crystals consisting of alternatively arranged dielectric and metallic layers. In Sec.2.1, we start with a one-dimensional photonic crystal (1DPC) of dielectric-metal layers and show that the principal values of the effective permittivity tensor along the two directions parallel to the material layers is of the same form as in the Drude model, implying a plasma behavior, whereas in the direction perpendicular to the layers, the effective medium mimics a Lorentz medium. We then derive the formula of group velocity for propagating mode in Sec. II.B and discuss in Sec. II.C the restrictions on its propagation directions and prove rigorously in Sec. II.D that they must be slower than the light speed in vacuum if the dielectric materials used in this 1DPC is a real material having relative permittivity larger than one. We also derive the formula of the dynamical wave energy density in Sec.II.E for arbitrary time varying electromagnetic fields in the effective medium. The energy formula is then used in the numerical study of the energy propagation velocity of a wave packet in Sec. III.A and III.B The wave packet is formed by a supposition of many plane waves with properly chosen coefficients for different propagation directions and frequencies. We then compare the numerical results for the wave packet and the theoretical results for the propagating mode and discuss the reasons of their differences. The conclusion is presented in Sec.IV.

## II. Model system and theoretical analysis

### A. Effective permittivity for the layered structure

To begin with, we consider a one-dimensional periodic structure which has one metal layer and one dielectric layer per unit cell, as is shown in Fig.1. In the long wavelength limit this photonic crystal behaves like an effective anisotropic medium which in some frequency ranges has indefinite eigenvalues of its permittivity tensor. We arrange these layers to be parallel to the *xy* plane and periodic along the *z* axis, so if we define *xz* plane to be the plane of incidence then the electromagnetic (EM) fields in this structure can be set to be independent of the *y* coordinate. Since the anisotropic feature appears only for the TM ($E_x$ $E_z$ $H_y$) polarized waves, hereafter we consider only waves of this polarization.

Denoting the filling fraction of the dielectric layer in one unit cell as *F*, and $\varepsilon_d$ and $\varepsilon_m$ the permittivity of the dielectric and metallic layer respectively, the principal values of the effective permittivity tensor of the metamaterial along the *x* and *z* directions under the long wavelength limit are given by:

$$\varepsilon_x = F\varepsilon_d + (1-F)\varepsilon_m$$
$$\varepsilon_z = \frac{\varepsilon_d \varepsilon_m}{F\varepsilon_m + (1-F)\varepsilon_d} \quad (1)$$

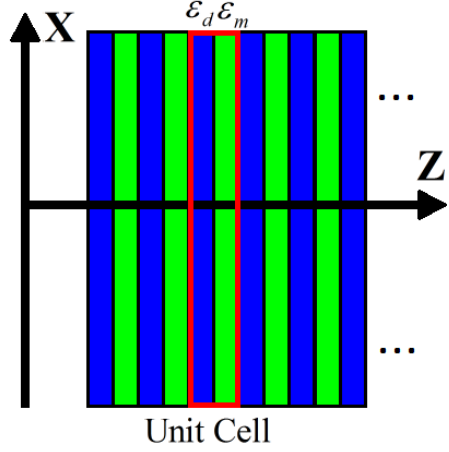

FIG. 1. The layered structure consisting of metal (green) and dielectric (blue) layers. The red rectangle represents a unit cell of the structure

We assume that the permittivity $\varepsilon_d$ of the dielectric layers is a constant and the metal permittivity $\varepsilon_m$ is described by the Drude model with negligible dissipation effect, that is:

$$\varepsilon_m = \varepsilon_0 \left(1 - \frac{\omega_p^2}{\omega^2}\right) \tag{2}$$

where $\varepsilon_0$ is the dielectric constant of the empty space.

The symbol $\omega_p$ stands for the plasma frequency which depends on the mass and concentration (number density) of the electrons in the metal. Substituting Eq. (2) to Eq. (1), we get

$$\varepsilon_x = \varepsilon_0 A_x \left(\frac{\omega^2 - \omega_p^{eff\,2}}{\omega^2}\right) \equiv \varepsilon_0 \tilde{\varepsilon}_x$$
$$\varepsilon_z = \varepsilon_0 A_z \left(\frac{\omega^2 - \omega_p^2}{\omega^2 - \omega_0^2}\right) \equiv \varepsilon_0 \tilde{\varepsilon}_z \tag{3}$$

where $\tilde{\varepsilon}_x$ and $\tilde{\varepsilon}_z$ are the principal values of the relative permittivity along the x and z axes. The coefficients $A_x$ and $A_z$ are given in Eq.(4). We observe that under the long wavelength limit, the effective permittivity in the *x* direction still has the Dude-model form, whereas along the *z* direction the metamaterial behaves like a polaritonic medium having Lorentz-type dispersion. That is, along the *z* direction there is a frequency gap between two resonance frequencies $\omega_0$ and $\omega_p$. Hereafter we denote the relative permittivity of the dielectric layers as $\tilde{\varepsilon}_d = \varepsilon_d / \varepsilon_0$. The parameters in Eq. (3) are given by:

$$A_x = (1-F) + F\tilde{\varepsilon}_d$$

$$A_z = \frac{\tilde{\varepsilon}_d}{\left[(1-F)\tilde{\varepsilon}_d + F\right]}$$

$$\omega_p^{eff} = \omega_p \sqrt{\frac{(1-F)}{F\tilde{\varepsilon}_d + (1-F)}}$$

$$\omega_0 = \omega_p \sqrt{\frac{F}{(1-F)\tilde{\varepsilon}_d + F}}$$

(4)

Here both $\omega_p^{eff}$ and $\omega_0$ are functions of $F$, and they become the same if $F = 1/2$.

## B. Group velocity of a propagating mode in a dispersive anisotropic medium

In this section we derive the formula for the group velocity of a propagating mode in a general anisotropic dispersive medium. This velocity can be further shown to be equal to the energy velocity defined by the ratio between the time averaged Poynting vector and the time averaged energy density. We will not derive this result here but only recognize it as a fact.

The dispersion relation for TM propagating mode in an anisotropic dispersive medium characterized by the principal values of permittivity $\varepsilon_z = \varepsilon_0 \tilde{\varepsilon}_z$ and $\varepsilon_x = \varepsilon_0 \tilde{\varepsilon}_x$ can be derived from Maxwell's equations as

$$\frac{k_x^2}{\tilde{\varepsilon}_z} + \frac{k_z^2}{\tilde{\varepsilon}_x} = \frac{\omega^2}{c^2} \quad (5)$$

The group velocity $\mathbf{v_g} = \nabla_k \omega$ for this mode can be deduced by taking the k-gradient of the dispersion relation:

$$\frac{2}{\tilde{\varepsilon}_z} k_x \hat{x} + \frac{2}{\tilde{\varepsilon}_x} k_z \hat{z} - \left(\frac{k_z}{\tilde{\varepsilon}_x}\right)^2 \nabla_k \tilde{\varepsilon}_x - \left(\frac{k_x}{\tilde{\varepsilon}_z}\right)^2 \nabla_k \tilde{\varepsilon}_z = \frac{2\omega}{c^2} \nabla_k \omega \quad (6)$$

Since the principal values of the permittivity of the medium are frequency dependent, so there are the relations $\nabla_k \tilde{\varepsilon}_x = \frac{\partial \tilde{\varepsilon}_x}{\partial \omega} \nabla_k \omega$ and $\nabla_k \tilde{\varepsilon}_z = \frac{\partial \tilde{\varepsilon}_z}{\partial \omega} \nabla_k \omega$. Substituting them into Eq.(6), we get

$$\frac{2}{\tilde{\varepsilon}_z} k_x \hat{x} + \frac{2}{\tilde{\varepsilon}_x} k_z \hat{z} = \left[\frac{2\omega}{c^2} + \left(\frac{k_z}{\tilde{\varepsilon}_x}\right)^2 \frac{\partial \tilde{\varepsilon}_x}{\partial \omega} + \left(\frac{k_x}{\tilde{\varepsilon}_z}\right)^2 \frac{\partial \tilde{\varepsilon}_z}{\partial \omega}\right] \mathbf{v_g} \quad (7)$$

Multiplying Eq.(7) by $\omega$ and using the dispersion relation (5), we have

$$2\omega \left(\frac{k_x}{\tilde{\varepsilon}_z} \hat{x} + \frac{k_z}{\tilde{\varepsilon}_x} \hat{z}\right) = \frac{1}{\omega} \left[\left(\frac{k_z}{\tilde{\varepsilon}_x}\right)^2 \frac{\partial (\omega^2 \tilde{\varepsilon}_x)}{\partial \omega} + \left(\frac{k_x}{\tilde{\varepsilon}_z}\right)^2 \frac{\partial (\omega^2 \tilde{\varepsilon}_z)}{\partial \omega}\right] \mathbf{v_g} \quad (8)$$

thus the group velocity in the dispersive anisotropic medium is:

$$\mathbf{v_g} = \frac{2\omega^2 \left(\frac{k_x}{\tilde{\varepsilon}_z} \hat{x} + \frac{k_z}{\tilde{\varepsilon}_x} \hat{z}\right)}{\left(\frac{k_z}{\tilde{\varepsilon}_x}\right)^2 \frac{\partial (\omega^2 \tilde{\varepsilon}_x)}{\partial \omega} + \left(\frac{k_x}{\tilde{\varepsilon}_z}\right)^2 \frac{\partial (\omega^2 \tilde{\varepsilon}_z)}{\partial \omega}} \quad (9)$$

## C. The restrictions on the propagating directions of the modes

In this section we discuss the existence condition and the restrictions on the propagation direction of a propagating mode. To be more specific, we define

$$k_x = k \sin \theta, \quad k_z = k \cos \theta, \quad k = \sqrt{k_x^2 + k_z^2} \tag{10}$$

For a propagating mode all of the three quantities must be real, and conventionally $k$ takes positive value. Note that according to Eq.(5) if both $\tilde{\varepsilon}_x$ and $\tilde{\varepsilon}_z$ were positive there would be no restriction on the propagating direction. On the other hand, if both of them were negative there would be no propagating mode at all. For the interested indefinite situation $\tilde{\varepsilon}_x \tilde{\varepsilon}_z < 0$, we have either $\tilde{\varepsilon}_x < 0$ or $\tilde{\varepsilon}_z < 0$. Substituting Eq.(10) into Eq.(5), we get

$$\left( \frac{1}{\tilde{\varepsilon}_x} - \frac{1}{\tilde{\varepsilon}_z} \right) \sin^2 \theta = \frac{1}{\tilde{\varepsilon}_x} - \frac{\omega^2}{k^2 c^2} < \frac{1}{\tilde{\varepsilon}_x} \tag{11}$$

Suppose we have $\tilde{\varepsilon}_x > 0 > \tilde{\varepsilon}_z$, then Eq.(11) implies that the propagating mode can only have wave vector oriented inside the angular region $|\theta| < \sin^{-1} \frac{1}{\sqrt{1+|\tilde{\varepsilon}_x/\tilde{\varepsilon}_z|}}$. Similarly, if we have $\tilde{\varepsilon}_z > 0 > \tilde{\varepsilon}_x$, then the wave vector must oriented in the region $|\theta| > \sin^{-1} \frac{1}{\sqrt{1+|\tilde{\varepsilon}_x/\tilde{\varepsilon}_z|}}$.

The signs of the principal values of permittivity defined in Eq.(3) are controlled by the wave frequency $\omega$ and the values of $\omega_0$, $\omega_p$, and $\omega_p^{eff}$, and these characteristic frequencies are determined by the filling fraction $F$ through Eq.(4). For simplicity we define "boundary angle" as $\theta_b = \sin^{-1} \frac{1}{\sqrt{1+|\tilde{\varepsilon}_x/\tilde{\varepsilon}_z|}}$ and assume $\tilde{\varepsilon}_d = \varepsilon_d / \varepsilon_0 > 1$. For $F$ belonging to different ranges of frequency we have the following three tables:

**1.** $0 < F < 0.5 \Rightarrow \omega_0 < \omega_p^{eff} < \omega_p$

| $0 < F < 0.5$ | $0 \sim \omega_0$ | $\omega_0 \sim \omega_p^{eff}$ | $\omega_p^{eff} \sim \omega_p$ | $\omega_p \sim \infty$ |
|---|---|---|---|---|
| $\tilde{\varepsilon}_x$ | − | − | + | + |
| $\tilde{\varepsilon}_z$ | + | − | − | + |
| $|\theta|$ | $|\theta| > \theta_b$ | No solution | $|\theta| < \theta_b$ | No restriction |

**2.** $0.5 < F < 1 \Rightarrow \omega_p^{eff} < \omega_0 < \omega_p$

| $\frac{1}{2} < F < 1$ | $0 \sim \omega_p^{eff}$ | $\omega_p^{eff} \sim \omega_0$ | $\omega_0 \sim \omega_p$ | $\omega_p \sim \infty$ |
|---|---|---|---|---|
| $\tilde{\varepsilon}_x$ | $-$ | $+$ | $+$ | $+$ |
| $\tilde{\varepsilon}_z$ | $+$ | $+$ | $-$ | $+$ |
| $|\theta|$ | $|\theta| > \theta_b$ | No restriction | $|\theta| < \theta_b$ | No restriction |

**3.** $F = \frac{1}{2} \Rightarrow \omega_0 = \omega_p^{eff} < \omega_p$

| $F = \frac{1}{2}$ | $0 \sim \omega_0$ | $\omega_0 \sim \omega_p$ | $\omega_p \sim \infty$ |
|---|---|---|---|
| $\tilde{\varepsilon}_x$ | $-$ | $+$ | $+$ |
| $\tilde{\varepsilon}_z$ | $+$ | $-$ | $+$ |
| $|\theta|$ | $|\theta| > \theta_b$ | $|\theta| < \theta_b$ | No restriction |

These tables summarize the existence conditions for propagating modes and the restrictions on their propagating directions. The $F = 0$ and $F = 1$ cases respectively correspond to pure metallic and pure dielectric materials. In the next section we will show that if $\tilde{\varepsilon}_d = \varepsilon_d / \varepsilon_0 > 1$ were assumed, any propagating mode can only have group velocity slower than the speed of light in vacuum.

### D. The subluminality of the propagating modes

To discuss the speed of a propagating mode, we define a factor $\beta_\mathbf{g} = |\mathbf{v}_g|/c$, which represents the ratio between the group speed and the universal speed of light in vacuum. If $\beta_\mathbf{g}$ can be proved to be always smaller than 1, then the subluminality is established. According to Eq.(9) and Eq.(10), $\beta_\mathbf{g}$ takes the value:

$$\beta_g = \frac{|\mathbf{v}_g|}{c} = \frac{\sqrt{\left(\frac{\sin^2\theta}{\tilde{\varepsilon}_z} + \frac{\cos^2\theta}{\tilde{\varepsilon}_x}\right)\left[\left(\frac{\sin\theta}{\tilde{\varepsilon}_z}\right)^2 + \left(\frac{\cos\theta}{\tilde{\varepsilon}_x}\right)^2\right]}}{\left|\frac{(\omega^2\tilde{\varepsilon}_z)'}{2\omega}\left(\frac{\sin\theta}{\tilde{\varepsilon}_z}\right)^2 + \frac{(\omega^2\tilde{\varepsilon}_x)'}{2\omega}\left(\frac{\cos\theta}{\tilde{\varepsilon}_x}\right)^2\right|} \quad (12)$$

Which can be rewritten as $\beta_g = \sqrt{\gamma\eta}$, where

$$\gamma = \frac{\left(\frac{k_x}{\tilde{\varepsilon}_z}\right)^2 + \left(\frac{k_z}{\tilde{\varepsilon}_x}\right)^2}{\frac{(\omega^2 \tilde{\varepsilon}_z)'}{2\omega}\left(\frac{k_x}{\tilde{\varepsilon}_z}\right)^2 + \frac{(\omega^2 \tilde{\varepsilon}_x)'}{2\omega}\left(\frac{k_z}{\tilde{\varepsilon}_x}\right)^2}, \quad \eta = \frac{\tilde{\varepsilon}_z\left(\frac{k_x}{\tilde{\varepsilon}_z}\right)^2 + \tilde{\varepsilon}_x\left(\frac{k_z}{\tilde{\varepsilon}_x}\right)^2}{\frac{(\omega^2 \tilde{\varepsilon}_z)'}{2\omega}\left(\frac{k_x}{\tilde{\varepsilon}_z}\right)^2 + \frac{(\omega^2 \tilde{\varepsilon}_x)'}{2\omega}\left(\frac{k_z}{\tilde{\varepsilon}_x}\right)^2} \qquad (13)$$

If both $\gamma$ and $\eta$ can be proved to be smaller than 1, than subluminality must be true. Using Eq.(3), we find:

$$\frac{(\omega^2 \tilde{\varepsilon}_x)'}{2\omega} = A_x, \quad \frac{(\omega^2 \tilde{\varepsilon}_z)'}{2\omega} = A_z\left[1 + \frac{\omega_0^2(\omega_p^2 - \omega_0^2)}{(\omega^2 - \omega_0^2)^2}\right] \qquad (14)$$

Note that according to Eq.(4), if we assume $\tilde{\varepsilon}_d = \varepsilon_d / \varepsilon_0 > 1$ then we have $A_x > 1$ and $A_z > 1$, and therefore the two factors in Eq.(14) are also larger than 1. This result directly leads to $\gamma < 1$ because its denominator is larger than the numerator.

For the $\eta$ factor, there are three situations we need to consider. First, if both $\tilde{\varepsilon}_x$ and $\tilde{\varepsilon}_z$ are positive, we have $\frac{(\omega^2 \tilde{\varepsilon}_x)'}{2\omega} > \tilde{\varepsilon}_x$ and $\frac{(\omega^2 \tilde{\varepsilon}_z)'}{2\omega} > \tilde{\varepsilon}_z$, and thus $\eta < 1$. In the second situation, we have $\tilde{\varepsilon}_x > 0 > \tilde{\varepsilon}_z$ and a positive numerator of $\eta$. Now if we replace the original numerator with the positive term $\tilde{\varepsilon}_x\left(\frac{k_z}{\tilde{\varepsilon}_x}\right)^2$, we still get a ratio smaller than one. Therefore we know in this situation we also have $\eta < 1$. The third situation is $\tilde{\varepsilon}_z > 0 > \tilde{\varepsilon}_x$. We can just exchange the role of $\tilde{\varepsilon}_x$ and $\tilde{\varepsilon}_z$ in the second situation and follow the same reasoning to get $\eta < 1$. There is no need to consider the case that both $\tilde{\varepsilon}_x$ and $\tilde{\varepsilon}_z$ are negative, because under this situation there is no propagating mode. Since for all the situations discussed above we always have $\gamma < 1$ and $\eta < 1$, we conclude that $\beta_g = \sqrt{\gamma\eta} < 1$ and the subluminality for any propagating mode is established. Note that in the above discussion we always assume the condition $\tilde{\varepsilon}_d = \varepsilon_d / \varepsilon_0 > 1$. Without this assumption, we would not be able to establish the subluminality in general. We thus get the main result of this paper: for the effective medium derived from the one-dimensional photonic crystal consisting of dielectric-metallic unit cell, if the dielectric material is a real material having relative permittivity larger than 1, then the mode velocity in this effective medium must be slower than the speed of light in vacuum. Numerical results for filling fraction $F = 0.25$, 0.4, 0.75 and 0.95 are shown in Fig.2. Here the "k-angle" denotes the angle (in degrees) between the wave vector **k** of the propagating mode and the z-axis (the direction of the periodicity), and the color in color bar represents the group velocity to light speed ratio $\beta_g$ defined in Eq.(12). The blank regions represent the forbidden combinations of $(\omega, \mathbf{k})$ for propagating modes. As one can see, all the results confirm the conclusion of subluminality.

Based on this discussion, now it's easy to understand that the superluminality of the mode or wave packet velocity in [27] is caused by the fact that the frequency dependence of the permittivity tensor were not taken into account in the derivation of the group velocity formula, equation (7).

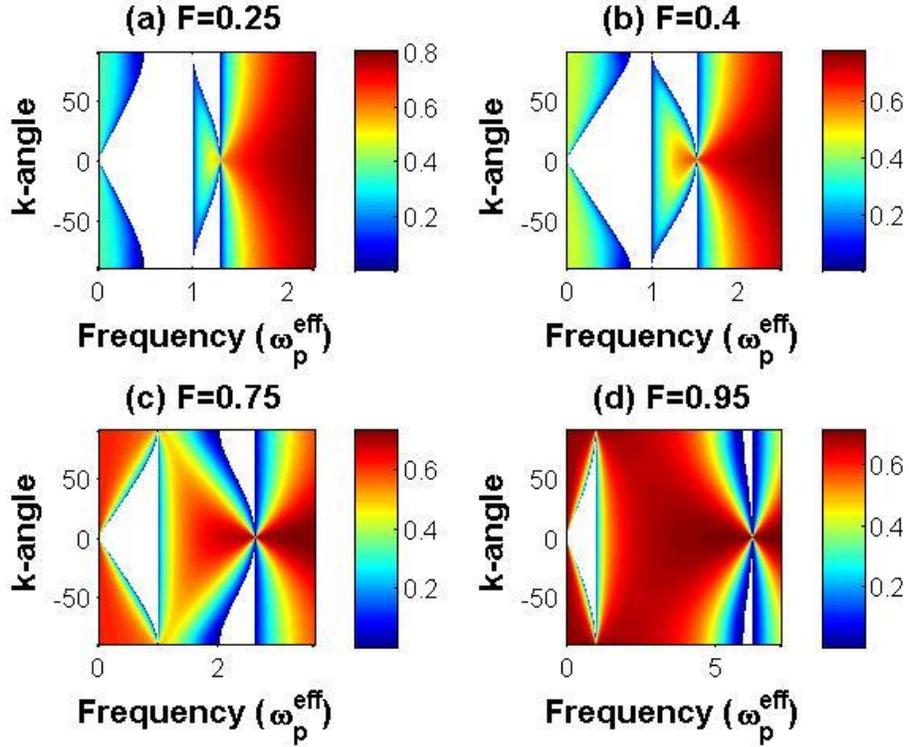

FIG. 2. The group velocity to light speed ratio for 4 filling fractions. The vertical axis represents the angle of the wave vector k for the mode. The unit frequency is the effective plasma frequency in Eq.(4). The permittivity of the dielectric in these simulation is 2. Every result shown here corresponds to subluminal group velocity.

**E. The energy density in the dispersive anisotropic metamaterial**

In the previous subsection, we have rigorously proved that the group velocity of a propagating mode in the effective hyperbolic metamaterial we defined in II.A must be subluminal. It is well known that for a propagating mode of frequency $\omega$ one can define its energy velocity as $\mathbf{v}_e = \dfrac{\langle \mathbf{S} \rangle}{\langle U \rangle}$ and this velocity can be proved to be equal to the group velocity, i.e, $\mathbf{v}_e = \mathbf{v}_g$ if the absorption effect can be neglected. Here $\langle A \rangle$ stands for the time average of the time-dependent quantity $A(t)$, $\mathbf{S} = \mathbf{E} \times \mathbf{H}$ is the Poynting vector, and $U$ is the energy density of the electromagnetic wave in the medium. However, instead of proving this relation, in the following sections we will compare $\mathbf{v}_g$ with the traveling speed of a wave packet which has central frequency $\omega$ and formed by a weighted sum of many propagating modes. We thus derive the instantaneous energy density formula here and apply it in the following sections for studying the energy propagation problem.

In classical electrodynamics the displacement field $\mathbf{D}(\mathbf{r},t)$ is related to the electric field $\mathbf{E}(\mathbf{r},t)$ and electric polarization field $\mathbf{P}(\mathbf{r},t)$ through the relation:

$$\mathbf{D} = \varepsilon_0 \mathbf{E} + \mathbf{P} \tag{15}$$

For the purpose of deriving the energy density formula of the EM field in the present medium, we find this relation is not very convenient to use. Instead of using Eq. (15) directly, we define a $\mathbf{Q}(\mathbf{r},t)$ field through the relations:

$$\begin{aligned} Q_x &= D_x - A_x \varepsilon_0 E_x \\ Q_z &= D_z - A_z \varepsilon_0 E_z \end{aligned} \tag{16}$$

Now we derive the dynamical relations between the $\mathbf{Q}(\mathbf{r},t)$ field and the applied electric field. Multiply both sides of Eq.(3) with $-\omega^2 E_x$ and $-\omega^2 E_z$, apply Eq.(16) and replace $-\omega^2$ with $\partial^2/\partial t^2$, we get:

$$\ddot{Q}_x = A_x \varepsilon_0 \omega_p^{eff\,2} E_x, \quad \ddot{Q}_z + \omega_0^2 Q_z = A_z \varepsilon_0 \left(\omega_p^2 - \omega_0^2\right) E_z \tag{17}$$

In fact, the terms $A_x \varepsilon_0 E_x$ and $A_z \varepsilon_0 E_z$ in Eq.(16) represent the $D_x$ and $D_z$ at infinite frequency, as can be checked from Eq.(3). They correspond to the non-dispersive response of the effective medium to the applied electric field. On the other hand, the $\mathbf{Q}(\mathbf{r},t)$ field represents the dynamical polarization field (the dispersive $\mathbf{P}$ field) of the medium when applying a time-varying electric field. Note that in this section all the quantities mentioned such as the electric field or the $\mathbf{Q}$-fields etc. are real value fields, not phasors (complex quantities).

Substituting Eq.(17) into Maxwell's equations, a Poynting theorem equation can be obtained. The left-hand side of this equation is the negative divergence of the Poynting vector $-\nabla \cdot \mathbf{S} = -\nabla \cdot (\mathbf{E} \times \mathbf{H})$, whereas the right-hand side is $\mathbf{E} \cdot \frac{\partial \mathbf{D}}{\partial t} + \mathbf{H} \cdot \frac{\partial \mathbf{B}}{\partial t}$. Similar to the previous works [29-30], the right hand side of the equation can be identified as the time rate of change of the field energy plus the power loss. Since in this paper we assume the power loss of the medium is small enough to be negligible, the right-hand side can be written as $\frac{\partial U}{\partial t} = \frac{\partial U_e}{\partial t} + \frac{\partial U_b}{\partial t}$. Here $\frac{\partial U_e}{\partial t} = \mathbf{E} \cdot \frac{\partial \mathbf{D}}{\partial t}$ is the rate of change of the electric energy density including both the contributions from the electric field itself and the material response of the medium. Similarly, $\frac{\partial U_b}{\partial t} = \mathbf{H} \cdot \frac{\partial \mathbf{B}}{\partial t} = \frac{\partial}{\partial t}\left(\frac{1}{2}\mu_0 H^2\right)$ is the rate of change of the magnetic field energy density. Finally, $U$ represents the total energy density to be identified. It is obvious that the magnetic energy density is simply given by $U_b = \frac{1}{2}\mu_0 H^2$, but the electric part $U_e$ is not that simple and need to be identified.

According to Eq.(17), we have:

$$\frac{\partial U_e}{\partial t} = \frac{\partial}{\partial t}\left\{\begin{array}{l}\frac{\varepsilon_0}{2}\left(A_x E_x^2 + A_z E_z^2\right) + \frac{1}{2\varepsilon_0 A_x \omega_p^{eff\,2}}\left(\frac{\partial Q_x}{\partial t}\right)^2 \\ + \frac{1}{2\varepsilon_0 A_z \left(\omega_p^2 - \omega_0^2\right)}\left[\left(\frac{\partial Q_z}{\partial t}\right)^2 + \omega_0^2 Q_z^2\right]\end{array}\right\} \tag{18}$$

thus the electric field energy density with material response should be identified as:

$$U_e = \frac{\varepsilon_0}{2}\left(A_x E_x^2 + A_z E_z^2\right) + \frac{1}{2\varepsilon_0 A_x \omega_p^2}\left(\frac{\partial Q_x}{\partial t}\right)^2 \\ + \frac{1}{2\varepsilon_0 A_z \left(\omega_p^2 - \omega_0^2\right)}\left[\left(\frac{\partial Q_z}{\partial t}\right)^2 + \omega_0^2 Q_z^2\right] \tag{19}$$

The electric energy density $U_e$ in Eq. (19) can be divided into two parts which can be recognized as the contributions from the electric field and the non-dispersive response of the medium (the first two terms on the right hand side of Eq.(19)) and the dispersive responses of the matter (the other terms). Once we know both the $\mathbf{E}$ and $\mathbf{H}$ fields in the medium, we can calculate $U_e$ from Eq.(19) and $U_b = \frac{1}{2}\mu_0 H^2$, thus the total energy density $U = U_e + U_b$ is obtained. The dispersive contribution (the second and third terms on the right hand side of Eq.(19)) of the energy density are from the kinetic and potential energies of the vibrating dipoles in the medium. The dynamical vibration of the dipoles is induced by the electric field in the medium.

In the next section we will numerically study the propagation of a wave packet in the hyperbolic medium. Two numerical velocities will be compared: the wave-amplitude velocity and the wave-energy velocity. The former is defined by tracking the motion of the amplitude center of the wave packet, whereas the latter is defined by tracking the motion of the energy-center of the wave packet, based on the formula derived in this subsection.

### III. Numerical simulation

### A. Building a wave packet based on the superposition principle

We can build a wave packet of arbitrary shape by adding the plane waves of propagating modes with appropriate weight function of frequency and wave vector. This is relying on the fact that the plane waves $e^{i(\mathbf{k}\cdot\mathbf{r}-\omega t)}$ with $\mathbf{k}$ and $\omega$ satisfying the dispersion relation Eq.(5) form a complete set, so any solution of Maxwell's equations in the medium can be written as a superposition of them. For numerical simulation of TM wave packet, we sum the magnetic fields and assume that the weight function has the following Gaussian form:

$$\mathbf{H_{y0}}\left(\omega, \mathbf{k}(\theta)\right) = H_0 \hat{y} \cdot e^{-\frac{(\omega-\omega_c)^2}{\omega_b^2}} \cdot e^{-\frac{(\theta-\theta_c)^2}{\theta_b^2}} \tag{20}$$

The $\omega_b$ and $\theta_b$ determine the distribution width of the plane wave bases in frequency and in propagating direction, respectively, whereas the $\omega_c$ and $\theta_c$ represent the central frequency and the central propagating direction (hereafter we call it incidence angle) of the wave vectors. The wave packet consisting of $m$ frequencies and $n$ wave vectors is thus written as:

$$\mathbf{H_y}(x,z;t) = \sum_{\omega=\omega_1}^{\omega_m} \sum_{k=\vec{k}_1}^{\vec{k}_n} \mathbf{H_{y0}}(\omega,\mathbf{k}) \cdot e^{i(\mathbf{k}\cdot\mathbf{r}-\omega t)} \tag{21}$$

In practice, the number of wave bases for constructing the wave packet should be large enough to prevent the possibility of more than one packets appear in our interest spatial region of observation.

### B. The propagation of wave packet in a hyperbolic metamaterial

As mentioned before, the signs of $\tilde{\varepsilon}_x$ and $\tilde{\varepsilon}_z$ are determined by the wave frequency $\omega$ and three characteristic frequencies $\omega_0$, $\omega_p$, and $\omega_p^{eff}$, and these parameters are determined by the filling fraction $F$ through Eq.(4), as summarized in tables A to C. According to these results, the propagating directions of the modes are restricted by the condition $|\theta| > \theta_b$ or $|\theta| < \theta_b$, or without any restriction. To simplify our discussion without reducing relevant generality, we assume $F = \frac{1}{2}$, i.e., the case in Table C. Hereafter we also assume $\tilde{\varepsilon}_d = 2$. Under this choice the $\tilde{\varepsilon}_x$ and $\tilde{\varepsilon}_z$ as functions of frequency are shown in Fig. 3.

Notice that in Fig.3, there are three different frequency bands characterized by the signs of the permittivity pair $(\tilde{\varepsilon}_x, \tilde{\varepsilon}_z)$: $(-,+)$, $(+,-)$, and $(+,+)$. Similarly, the theoretical group velocity (Eq.(12)) as function of the phase propagating angle (the wave vector direction of the mode) and frequency also exhibits three different patterns, as shown in Fig.4. The similar behaviors of the permittivity tensor and group velocity reflect the dispersive and anisotropic features of the effective medium.

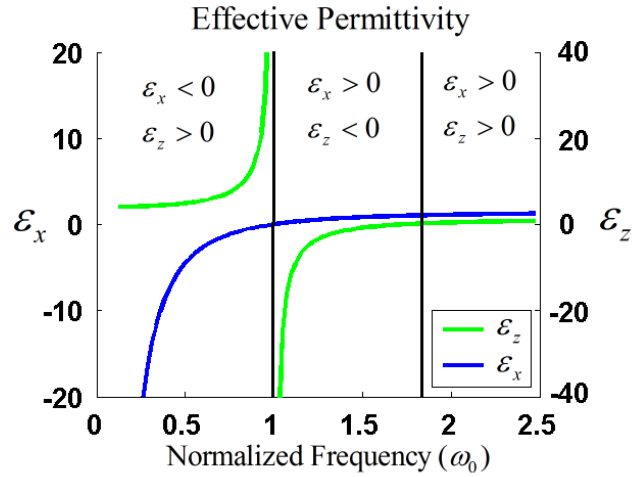

FIG. 3. The permittivity curves as functions of frequency. Here $\omega_0$ is chosen as the unit of frequency.

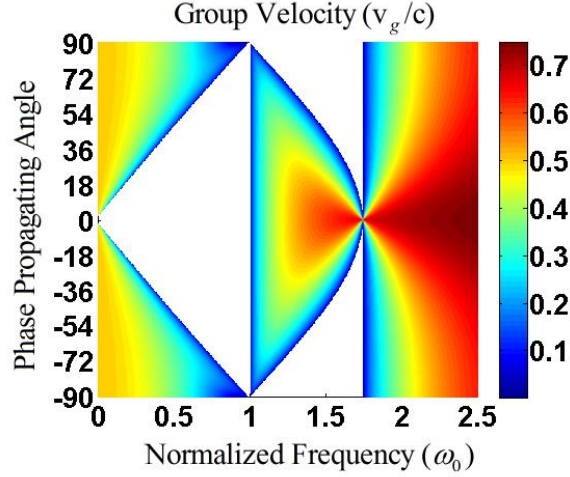

FIG. **4**. The theoretical group velocity of the propagating mode for filling fraction 0.5. The blank regions indicate that there is no propagating mode corresponding to such combination of direction and frequency. The phase propagating angle (in degrees) is the angle between the k-vector of the propagating mode and the z-axis.

In this research, we are not interested in the third region of Fig.3 because in it both of the two principal values the permittivity are positive. We will only consider the case of indefinite medium with hyperbolic dispersion relation. Since the first and second regions are not very different, we can choose either one of them. Hereafter we choose the second region as the frequency domain and do all the numerical simulations. Under this choice the effective medium has positive $\varepsilon_x$ and negative $\varepsilon_z$.

Now we can build the wave packet using Eq.(21). We define the wave packet center just like one defines the center of mass in a dynamical problem, but for the present problem the mass distribution is replaced by the magnitude of the instantaneous H-field or the energy density of the electromagnetic fields (including the response energy of the medium). We trace the motion of the center of the wave packet numerically, and the numerical "group velocity" of the wave packet is defined as the group center displacement in the time interval of one step of time counting. The typical field energy density distribution at three different times is shown in Fig.5. In this simulation we choose $\omega_c = 1.15\,\text{GHz}$, $\omega_b = 0.14\,\text{GHz}$, $\theta_b = 20°$, $m = n = 25$ (625 plane waves), and the energy density distribution for $t = 0, 15, 30\,\mu s$ are shown in the figure.

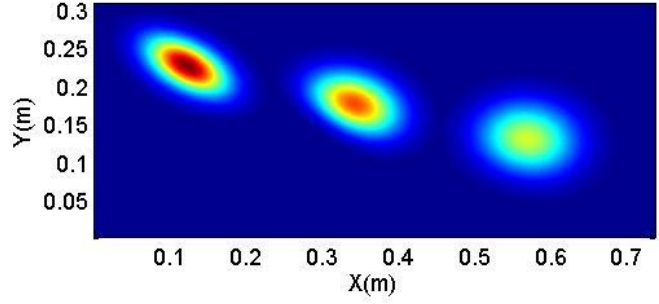

FIG. **5**. The distribution of the energy density at three different times.

The comparison between the mode group velocity (Eq.(9)) and the numerical wave packet velocities in the indefinite frequency domain for central phase propagating angle of 10 degrees and 30 degrees are shown as Fig 6. From it we find that for the wave-packet propagation, no matter whether we evaluate its velocity based on the amplitude or energy distribution, the result is very close to the theoretical group velocity of the mode if the bandwidth of the wave packet is not too wide. However, if the bandwidth of the wave packet is a little wider, the two possible packet velocities deviate a little bit from the theoretical group velocity of the mode. This is because some Fourier components the wave packet become evanescent waves and cannot penetrate through the media, and this leads to the shift of the central frequency of the wave packet.

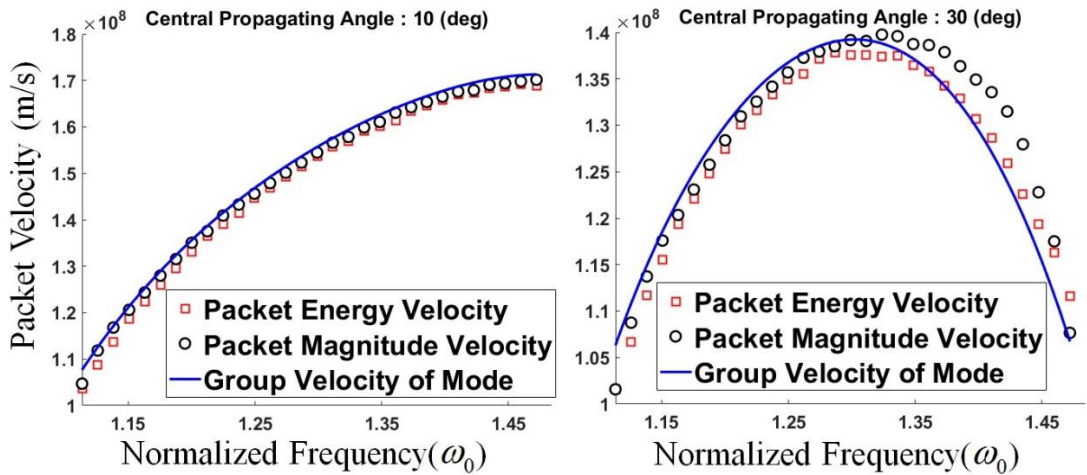

FIG. **6**. The comparison of the theoretical mode group velocity and the numerical wave packet velocities in the indefinite frequency domain for two central phase propagating angles.

In Fig. 7, we compare the velocity diagrams of the theoretical group velocity of the propagating modes and the numerical wave packet velocity defined with the energy density scheme. In constructing the numerical velocity diagram, various cases including 30 central frequencies and 20 central phase propagating angles are considered. The similarity between the theoretical and numerical results is very obvious.

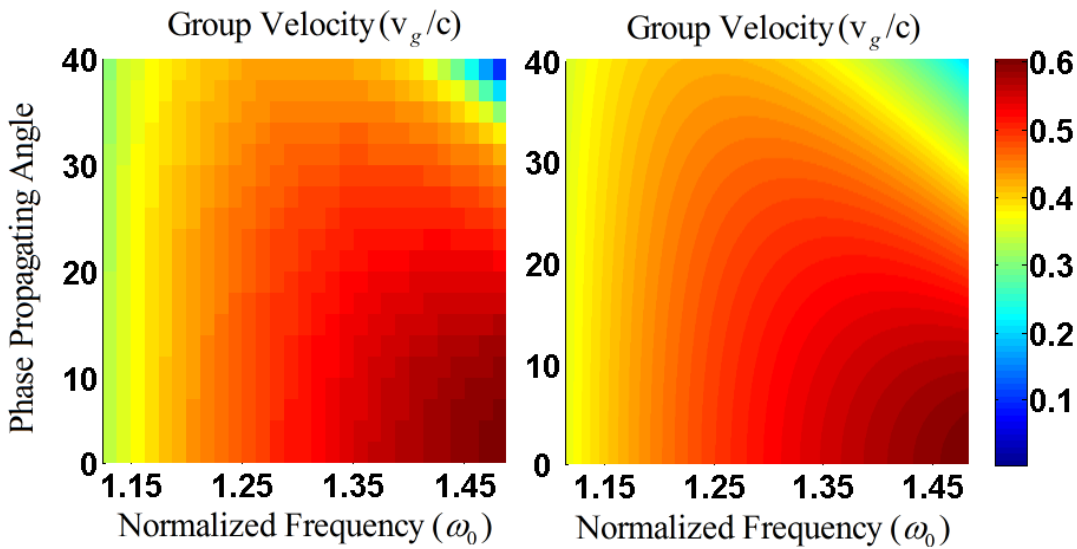

FIG. **7**. The distribution of the normalized group velocity in the indefinite frequency band with central phase propagation angle from 0 to 40 degrees

**IV. Conclusion**

In this paper, the effective electromagnetic properties in the long wavelength limit of the 1DPC of dielectric-metal layers is studied. This metamaterial behaves like a plasma or Lorentz medium with respect to the applied electric fields parallel or perpendicular to the layers, respectively. Based on the derivation of the Poynting theorem, the formula for the dynamical energy density is obtained. This dynamical energy includes two parts: the energies from the electric and magnetic fields themselves, and the kinetic plus the potential energies from the charge carriers in the medium. For the propagating modes in the medium, we derived the restrictions on the propagating directions of them. According to our theoretical analysis and numerical results, the group velocity of a propagating mode in the effective medium is always subluminal (slower than the speed of light) if the Drude and Lorentz dispersions already lead to subluminal propagation separately, that is, they are real materials. Thus our results confirm the conclusion of the theoretical work in [28] and reveal the reason of why superluminal propagation were derived in [27]. Besides, for the wave-packet velocity, no matter how we evaluate it, based on tracking the packet amplitude or energy distribution, the result is very close to the theoretical value of the group velocity for the propagating mode having the frequency close to the central frequency of the packet if the bandwidth of the packet is not too wide. If the bandwidth of the wave packet is a little wider, however, the two possible wave packet velocities (amplitude and energy) become deviated from the theoretical group velocity of the mode. We believe this is because some Fourier components of the wave packet become evanescent and cannot penetrate through the media, and this leads to the shift of the central frequency of the wave packet in the propagation process.

Our conclusion that no superluminal propagation of wave energy can happen in a dispersive-hyperbolic metamaterial is based on two main assumptions. The first is that the dielectric layers must have permittivity greater than 1, and the second is that the metal layers have permittivity of Drude-model type dispersion. Both of which are typical dielectric properties for real dielectric and metallic materials, respectively. It is in this sense we call them "real materials" in our paper. In addition, we assume through this paper that the medium is lossless. This assumption may seem unphysical because the Kramers-Kronig relations require a nonzero positive damping coefficient in the permittivity formula [31]. However, this requirement does not give a lower bound to such damping coefficient but only restricts the sign of it. Thus a lossless medium is not unphysical because it can be treated as an absorptive medium with a negligible damping coefficient. Finally, we want to stress that our conclusion does not exclude some superluminal phenomena happen in finite thickness gain medium or evanescent regions [32]. Our results are about wave propagation in an infinite or semi-infinite medium, which is very different from these superluminal phenomena caused by reshaping [16] or tunneling [18] effects in a finite region. Our results confirm that the wave energy propagation velocity in a real system cannot exceeds the speed of light in vacuum, which is a restriction derived from special relativity.